\documentclass[preprint,12pt]{elsarticle}
\usepackage{color,amsmath,amssymb,graphicx,latexsym,lineno}
\usepackage{ulem}
\usepackage[colorlinks=true]{hyperref}

\journal{Astroparticle Physics}

\begin{document}
\begin{frontmatter}



\title{The Influence of Sun's and Moon's Shadows on Cosmic-Ray Anisotropy}


\author[label1,label2]{Xuan'ang Ye}
\author[label1,label2]{Yi Zhang}
\author[label1,label2]{Jiayin He}
\author[label1,label2]{Shiping Zhao}
\affiliation[label1]{organization={Key Laboratory of Dark Matter and Space Astronomy \& Key Laboratory of Radio Astronomy, Purple Mountain Observatory, Chinese Academy of Science },
            city={Nanjing},
            postcode={210023},
            state={Jiangsu},
            country={China}}
\affiliation[label2]{organization={School of Astronomy and Space Science, University of Science and Technology of China},
            city={Hefei},
            postcode={230026},
            state={Anhui},
            country={China}}

\begin{abstract}
Large-scale anisotropy, with amplitudes reaching approximately 0.1\% at TeV energies, has been observed by multiple cosmic-ray experiments. The obstruction of cosmic rays by the Sun and Moon creates shadow effects, potentially impacting the observed cosmic ray anisotropy. To evaluate these effects, this study calculates the contributions of the Sun's and Moon's shadows to the overall cosmic-ray anisotropy in both local solar and sidereal time. The analysis reveals that in local sidereal time, the total 1D projection amplitude of the anisotropy is around 0.003\%, which is significantly smaller than the observed cosmic-ray anisotropy. This indicates that the influence of the Sun's and Moon's shadows on cosmic-ray anisotropy analysis in local sidereal time is negligible. In contrast, in local solar time, the shadow-induced deficit appears in a very small time bin, with a magnitude comparable to that of the cosmic-ray solar anisotropy. This deficit could serve as a benchmark for validating anisotropy measurements in future facilities.

\end{abstract}



\end{frontmatter}



\section{Introduction}
\label{label:introduction}
Cosmic rays (CRs) span energies from hundreds of GeV to PeV and are primarily of galactic origin. After propagating through the Milky Way's magnetic fields, the cosmic rays that reach Earth exhibit an almost isotropic directional distribution. However, a small anisotropy of approximately 0.1\% is observed at energies around the multi-TeV range \citep{Amenomori_aniso_2006, abdoMilagro_ANISOTROPY2009,agliettaEASTOP_ANISOTROPY2009, bartoliARGO_Anisotropy2018, abeysekaraHAWC_AnisotropyTeV2018,d.apelKasdadeAnisotropy2019, aartsenICECUBE_ANISOTROPY2016}. This anisotropy is largely attributed to the uneven distribution of CR sources, contributions from local sources, and the influence of regularities in the local magnetic field structure \cite{erlykinAnisotropy2006, harding2016explaining, ahlersDecipheringDipoleAnisotropy2016a}. Additionally, anisotropy can arise from the observer's motion relative to a reference frame with an isotropic cosmic ray distribution, known as the Compton–Getting effect\cite{comptonApparentEffectGalactic1935}, due to Earth's motion around the Sun. Precise measurements of these anisotropies can further enhance our understanding of cosmic-ray acceleration and propagation.

Numerous experiments have detected cosmic ray anisotropies in the energy range from approximately 500 GeV to 100 PeV. In the Northern Hemisphere, notable detectors include AS\(\gamma\) \cite{Amenomori_aniso_2006,asg_aniso_2017}, Milagro \cite{abdoMilagro_ANISOTROPY2009}, EAS-Top \cite{agliettaEASTOP_ANISOTROPY2009}, ARGO-YBJ \cite{bartoliARGO_Anisotropy2018}, HAWC \cite{abeysekaraHAWC_AnisotropyTeV2018}, KASCADE \cite{d.apelKasdadeAnisotropy2019}, GRAPES-3\cite{chakraborty2024small}, and LHAASO\cite{WCDA:anisotropy,KM2A:anisotropy}. In the Southern Hemisphere, key detectors are IceCube \cite{aartsenICECUBE_ANISOTROPY2016} and IceTop \cite{IceTop}. The dipole amplitude of the large-scale anisotropy typically hovers around 0.1\% in relative intensity, while middle-scale anisotropies range from 0.01\% to 0.1\%. At energies above EeV, significant measurements have only been made by the Pierre Auger Observatory, which observed a dipole amplitude of approximately 6.5\% \cite{augerAnisotropyScience2017}. 

The effects of the Moon and Sun on the deflection and obstruction of cosmic rays (CRs), leading to slight CR anisotropies, have not been extensively discussed and are generally overlooked in measurements. Nearby celestial bodies such as the Sun and Moon can obstruct some CRs, resulting in observable CR "shadows" on Earth. These shadows represent a reduction in the number of CRs coming from the direction of the Sun\cite{amenomori2013probe,refId0,PhysRevD.103.042005,Romanov_2021} or Moon\cite{PhysRevD.84.022003,PhysRevD.89.102004}. Traditionally, in CR anisotropy analysis, the impact of these shadows has been considered minimal and negligible. However, as the study of cosmic ray anisotropy has entered an era of precision measurement, with more CR events being recorded, the influence of the Sun's and Moon's shadows on CR anisotropy warrants further discussion.

In this paper, we perform latitude-dependent simulations of the Sun's and Moon's shadows in both local sidereal and solar time. We compare the anisotropy caused by these shadows with typical cosmic-ray anisotropies. The paper is organized as follows: Section 2 covers the simulation of the cosmic-ray background and the methods used for simulating the Sun's and Moon's shadows. Section 3 presents our analysis and results, while Section 4 provides discussion and conclusions.

\section{Method}
\label{label:method}

\begin{figure}
    \centering
    \includegraphics[width=0.5\textwidth]{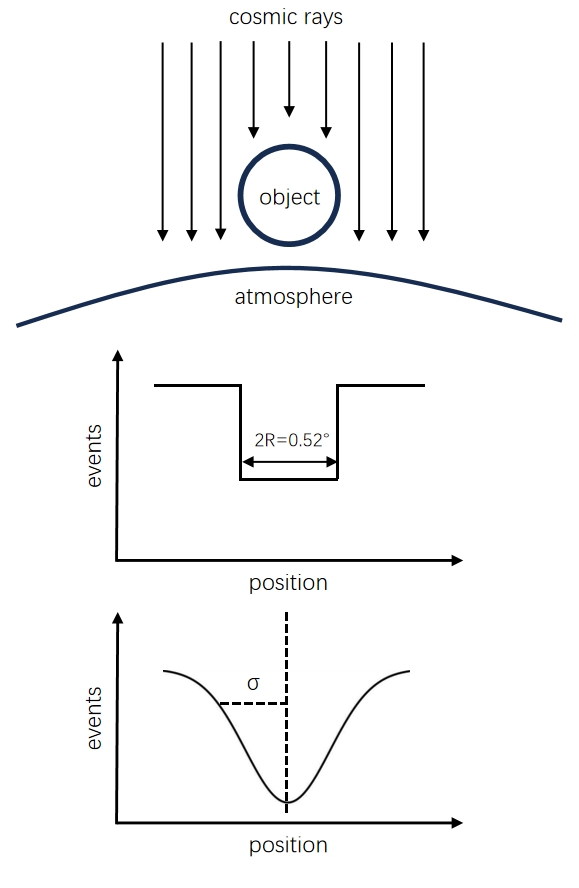}
    \caption{Schematic illustration of the formation of the cosmic-ray shadow of the Moon or Sun.}
    \label{fig:shadow_schematic}
\end{figure}

Fig. \ref{fig:shadow_schematic} illustrates the cosmic-ray deficit caused by the Sun and Moon. For simplicity, deflections of cosmic rays due to the interplanetary magnetic field (IMF) and the geomagnetic field (GMF) are not considered. The apparent angular radius R of both the Moon and the Sun is approximately $0.26^\circ$, within which cosmic rays are blocked by these celestial bodies. In the case of an ideal detector without angular resolution, the number of events within this circular area is zero, resulting in the observed deficit, as shown in the middle panel of Fig. \ref{fig:shadow_schematic}.

Due to limitations in the detector's angular resolution, the observed deficit shape does not precisely match the actual shapes of the Sun and Moon. Instead, it reflects a convolution of these shapes with the detector’s point spread function (PSF), as depicted in the third panel of Fig. \ref{fig:shadow_schematic}. If the detector's PSF is significantly larger than the Sun's apparent radius and follows an approximately Gaussian distribution, the observed deficit resembles a two-dimensional Gaussian distribution, forming the Sun's (or Moon's) shadow we observed.

Based on the above scenario, we simulated the cosmic-ray background, as well as the Sun's and Moon's shadows, to further investigate the effect of these shadows on cosmic-ray anisotropy. We generated this cosmic-ray background skymap in celestial coordinates (right ascension $\alpha$ and declination $\delta$), with a grid size of $0.1^\circ \times 0.1^\circ$. This simulation employed a typical spatial distribution of cosmic rays derived from ground-based extensive air shower experiments, assuming the site is located at a specified declination.

The spatial distribution of cosmic-ray events encompasses both the zenith angle $\theta$ distribution and the azimuth angle $\phi$ distribution. The zenith angle distribution of simulated cosmic-ray events follows the expression $A \cos^n \theta$, where $A$ represents the number of events detected per unit solid angle by the detector when $\theta=0$. The expression aligns with multiple experimental observations and is primarily attributed to the zenith-angle dependence of atmospheric depth\cite{grieder2010extensive}. For zenith angles less than $50^\circ$, the parameter $n$ is set to 4, with the maximum number of events occurring around a zenith angle of $29^\circ$. 
Here, $n$ is determined to be 4 based on results from toy Monte Carlo simulations. We generated cosmic-ray events using CORSIKA\cite{heck1998corsika} and simulated the detector response with scintillator detectors similar to those employed in LHAASO\cite{aharonianObservationCrabNebula2021}. The energy range considered was from 1 TeV to 1 PeV, with an initial zenith angle range of $0^\circ–70^\circ$, which was further narrowed to $0^\circ–50^\circ$ after applying cuts. The value of $n$ may vary slightly due to the specific experiment's altitude, latitude, and detector response.  However, we find that $n$ does not significantly affect the deficit ratio. Therefore, the discussion in this paper uses $n=4$ as an example. Additionally, the azimuth angle distribution is assumed to be uniform, primarily due to the consistent atmospheric depth within the same zenith angle band. Typically, within the same zenith angle band, the detection efficiency remains constant \cite{amenomoriNorthernSkySurvey2005}.


To simulate the shadows of the Sun and Moon, we utilize the EPHEM package \cite{rhodes2011pyephem} to determine the celestial coordinates, specifically the right ascension $\alpha$ and declination $\delta$, of celestial bodies such as the Moon or Sun. This allows us to obtain their zenith angle and azimuth angle $\phi$ for a specified time bin $t$. As illustrated in Fig. \ref{fig:shadow_schematic}, the distribution of the shadow deficit is represented by the convolution of a $0.26^\circ$ disk with the point spread function (PSF). In our analysis, the PSF is modeled as a two-dimensional Gaussian distribution with a width of $0.5^\circ$, which is consistent with the typical PSFs observed in ground-based experiments \cite{HAWC:2year-CRA,he2018design} at their threshold energy.

For a given time bin $t$, we simulate the expected cosmic-ray background as the $N_{off}^t$ map. We then calculate the expected number of deficits around the Sun (within a range of three times the Gaussian width) to create the $N_{signal}^t$ map. By subtracting the $N_{signal}^t$ map from the $N_{off}^t$ map, we obtain the $N_{on}^t$ map. We aggregate the $N_{off}^t$ and $N_{on}^t$ maps across all time bins to produce the $N_{off}$ and $N_{on}$ maps, respectively. Using these $N_{on}$ and $N_{off}$ maps, we define the relative intensity $I$ as
\begin{equation}
    I=\frac{N_{on}}{N_{off}}
\end{equation}
The deficit ratio is then calculated as $1-I$. This formula is employed to generate the deficit ratio map from the $N_{on}$ and $N_{off}$ maps.

\section{Results}
\label{label:results}

\subsection{Deficit Ratio in Local Sidereal Time}
\label{label:sid_deficit}

\begin{figure}
    \centering
    \includegraphics[width=\textwidth]{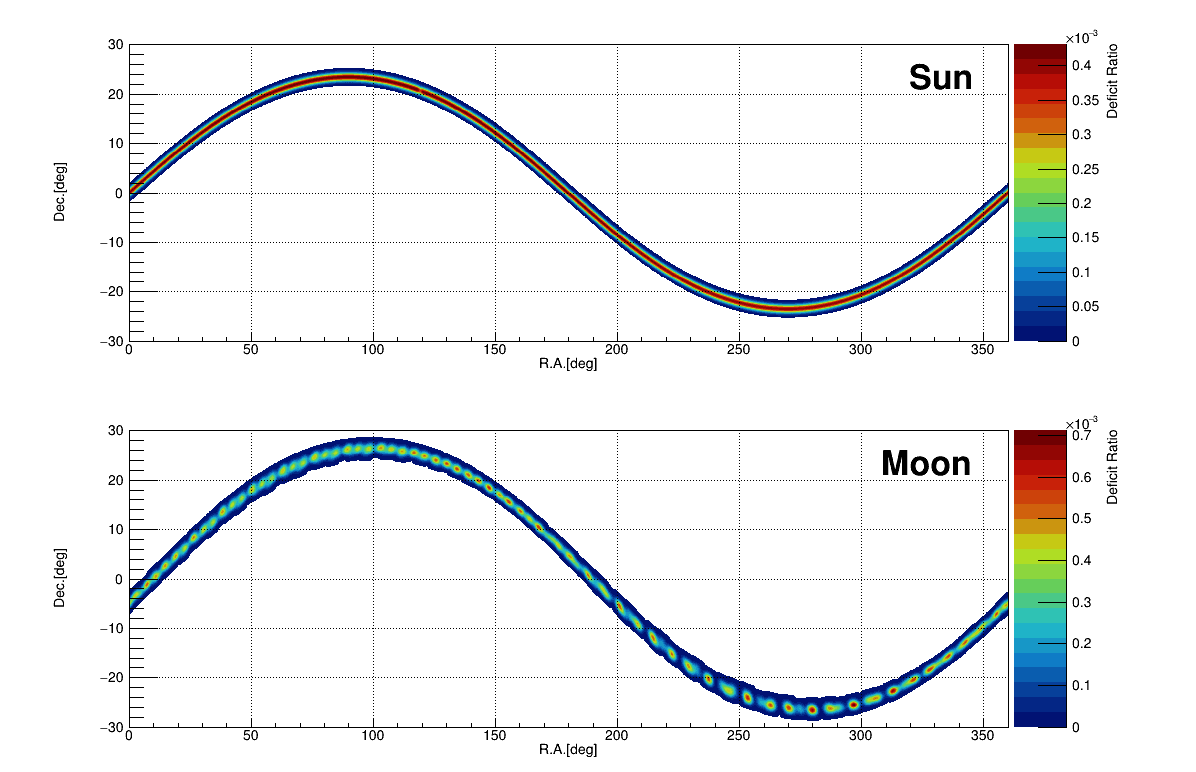}
     \caption{2D maps of the deficit ratio in local sidereal time for the Sun's shadow (top panel) and the Moon's shadow (bottom panel).}
    \label{fig:sid_2d}
\end{figure}

Fig. \ref{fig:sid_2d} presents the deficit ratio map for one year of observations, illustrating the shadows of the Sun (top) and Moon (bottom) as observed from a detector at a geographic latitude of $0^\circ$. Due to their orbital motion, the shadows of the Sun and Moon are confined to a narrow band in the sky throughout the year. The maximum deficit caused by the Sun is approximately $0.04\%$, while for the Moon it is about $0.07\%$. This difference in the deficit ratio reflects the varying duration each shadow occupies a specific grid cell. The deficit ratio is an annual average, and both the Sun and Moon occupy any given cell on the map for less than one day over the course of a year. For the Sun, its regular orbital path results in a relatively consistent deficit ratio across different regions of the sky. In contrast, the Moon's trajectory results in the formation of specific hotspots. This occurs because the Moon's sidereal month, approximately 27.3 days, covers 360 degrees of right ascension. The detector observes the Moon's shadow only at specific times each day, leading to a discontinuous trajectory. Since the orbital period is not an integer multiple of a day, the observed shadow shifts monthly, causing hotspots where shadows overlap.

\begin{figure}
    \centering
    \includegraphics[width=0.48\textwidth]{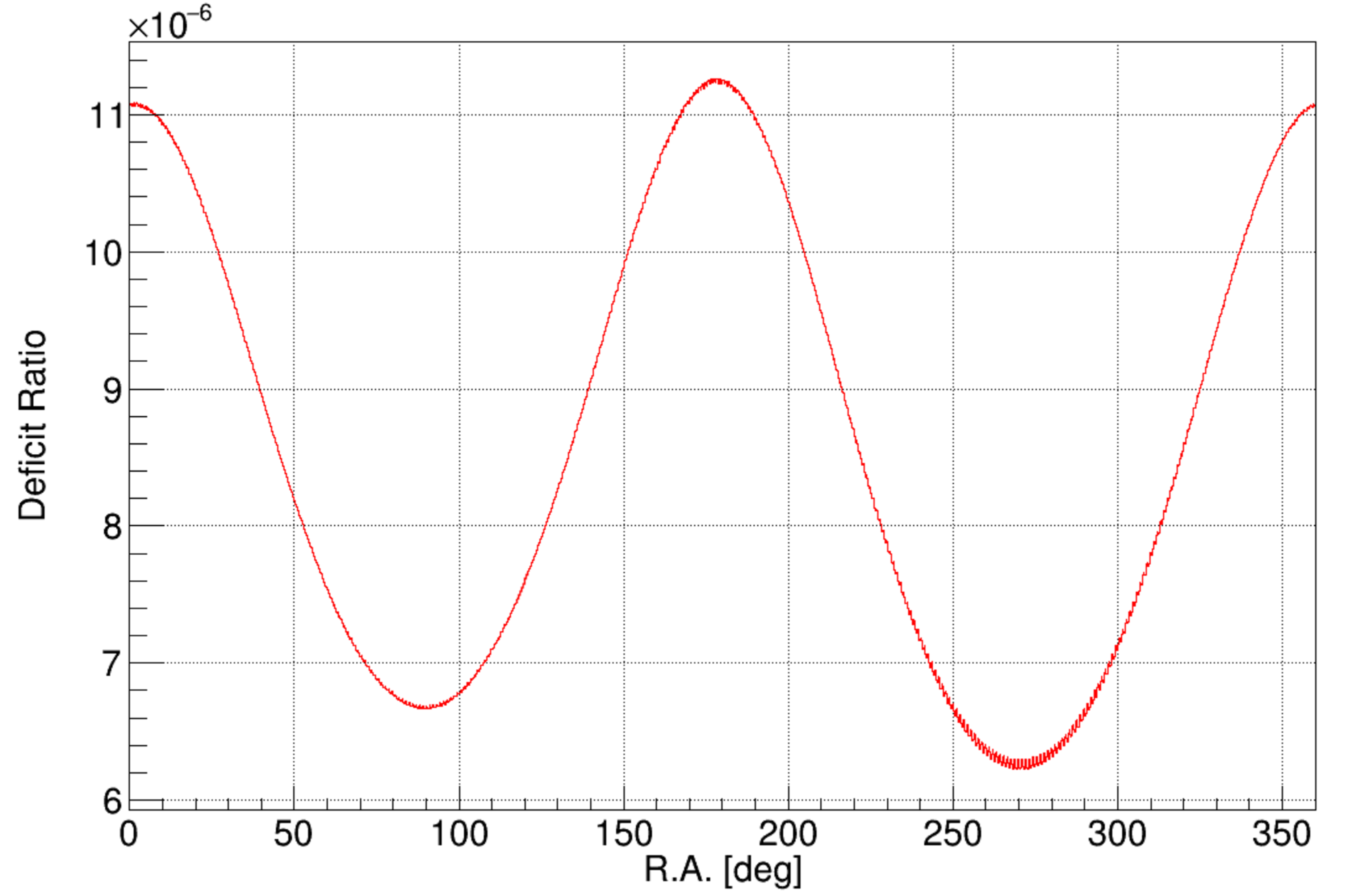}
    \includegraphics[width=0.48\textwidth]{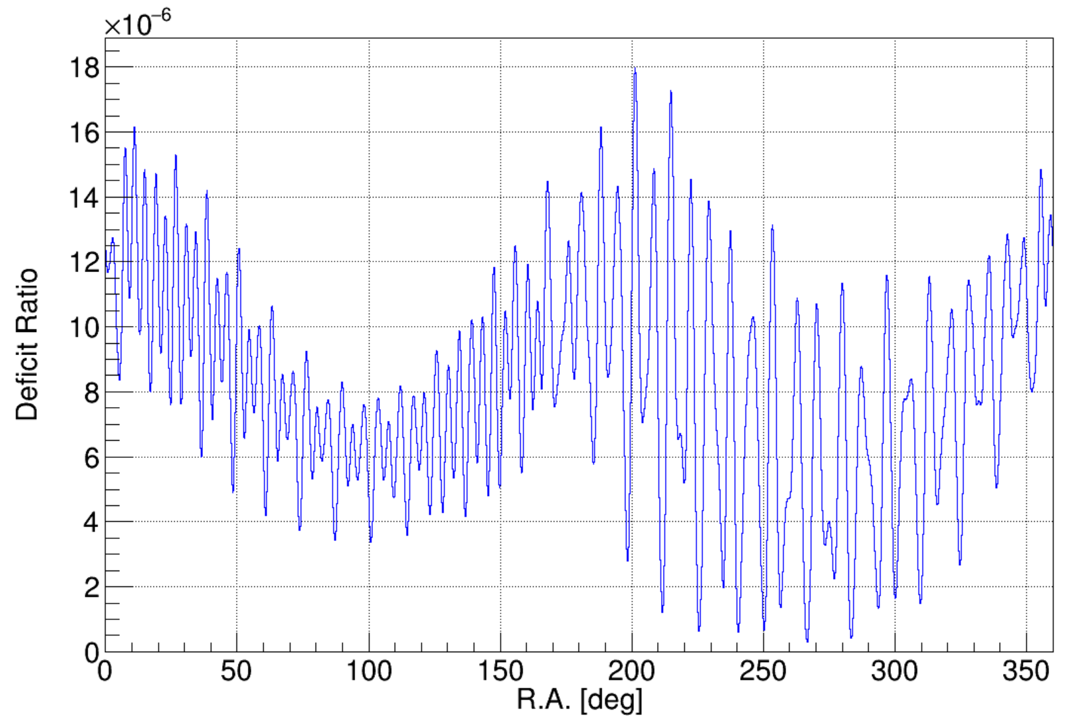}
   \caption{One-dimensional projection of the deficit ratio spanning a declination range from $-50^\circ$ to $50^\circ$ in local sidereal time for the Sun's shadow (left) and the Moon's shadow (right).}
    \label{fig:sid_1d}
\end{figure}

Fig. \ref{fig:sid_1d} presents a one-dimensional projection of the deficit ratio skymap along right ascension, spanning a declination range from -50° to 50°. In regions where the sky is blank, the relative intensity $I$ is 1, which reduces the overall average deficit ratio in the projection. At a given right ascension, the closer the Sun is to the detector’s geographic latitude, the smaller the zenith angle of the blocked cosmic-ray (CR) events, resulting in a higher projected deficit ratio due to more CR events being blocked. This trend is also observed for the Moon. Due to the Moon’s shorter orbital period around Earth, there are significant irregular fluctuations in its deficit ratio, as shown in Fig. \ref{fig:sid_1d}. The projected 1D deficit ratios for both the Sun and Moon peak when their paths align with the detector’s latitude of $0^\circ$, with the Sun’s maximum deficit ratio reaching 0.0011\% and the Moon’s reaching 0.0018\%.

\begin{figure}
    \centering
    \includegraphics[width=0.48\textwidth]{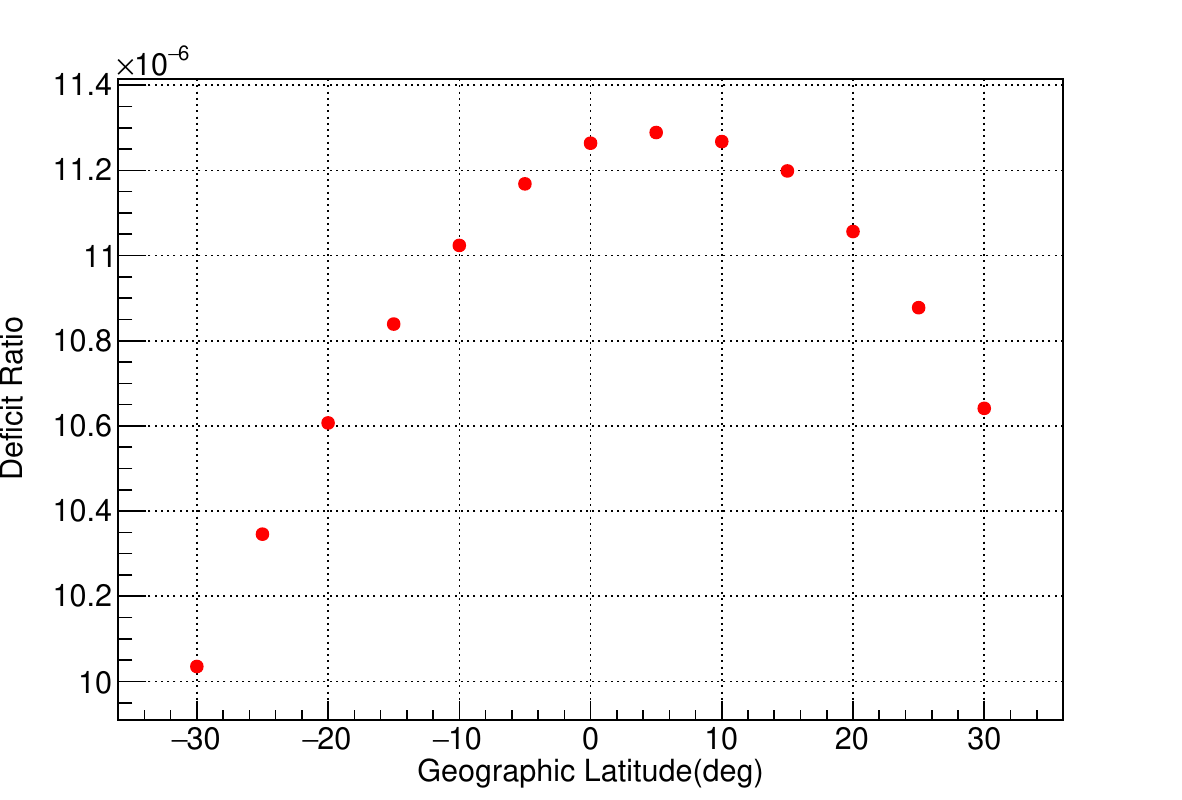}
    \includegraphics[width=0.48\textwidth]{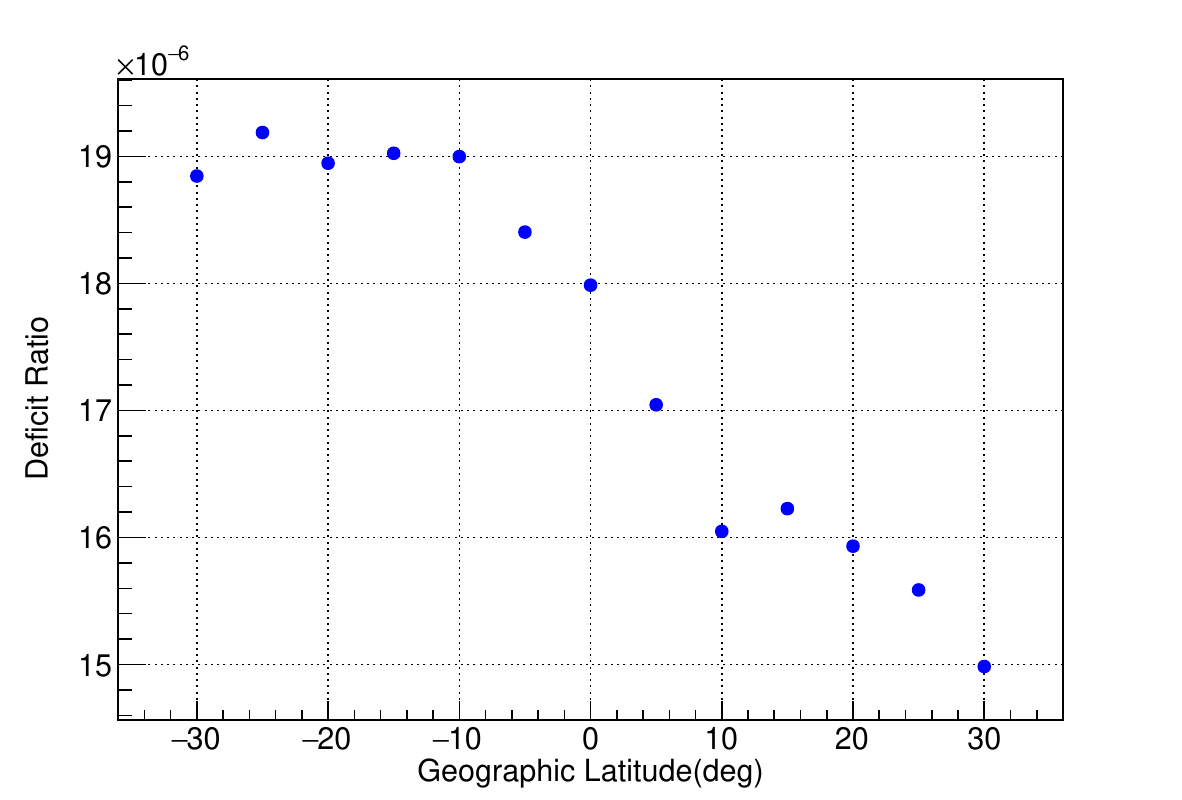}
    \caption{Maximum projected deficit ratio as a function of the detector's geographic latitude in local sidereal time, for the Sun's shadow (left) and the Moon's shadow (right).}
    \label{fig:sid_max_lat}
\end{figure}

Fig. \ref{fig:sid_max_lat} shows the relationship between the maximum 1D deficit ratios of the Sun's (left) and Moon's (right) shadows and the geographic latitude of the detector. The maximum deficit of the Sun's shadow varies as the detector moves within $\pm30^\circ$ geographic latitude, ranging from $1.128 \times 10^{-5}$ to \(1.006 \times 10^{-5}\); for the Moon, it ranges from \(1.919 \times 10^{-5}\) to \(1.498 \times 10^{-5}\). These levels are insufficient to significantly affect cosmic-ray anisotropy in local sidereal time, where typical amplitudes are around \(0.2\%\) at TeV energy (ARGO \cite{Chen_2019} and AS\(\gamma\) \cite{ASgamma}). This finding supports the conclusion that the influence of the Sun's and Moon's shadows on cosmic-ray anisotropy in local sidereal time is negligible in ground-based extensive air shower (EAS) experiments.

\subsection{Deficit Ratio in local Solar Time}
The anisotropy observed in local solar time primarily arises from the Compton-Getting (CG) effect, which results from Earth's orbital motion around the Sun. In this study, we calculate the Compton-Getting effect using the equation \cite{Gleeson1968}:
\begin{equation}
    \xi_{CG} = \beta(2+\gamma)\cos\theta
\end{equation}
Where \(\xi_{CG}\) represents the amplitude of anisotropy relative to the background cosmic-ray flux. The spectral index of cosmic rays, \(\gamma\), is commonly assumed to be 2.7 in the TeV energy range. The factor \(\beta = \frac{v}{c}\), where \(v\) is Earth's orbital speed and \(c\) is the speed of light. The angle \(\theta\) denotes the angle between Earth's orbital direction and the direction of the incident cosmic ray.

Taking the CG effect into account, $N_{on}^t$ for a given time bin is calculated using the formula: $N_{on}^t=\xi_{CG}(\theta)N_{off}^t-N_{signal}^t$. The $N_{on}$ maps, $N_{off}$ maps, relative intensity $I$, and the deficit ratio $1-I$ are subsequently determined using the same approach.

\begin{figure}
    \centering
    \includegraphics[width=0.5\textwidth]{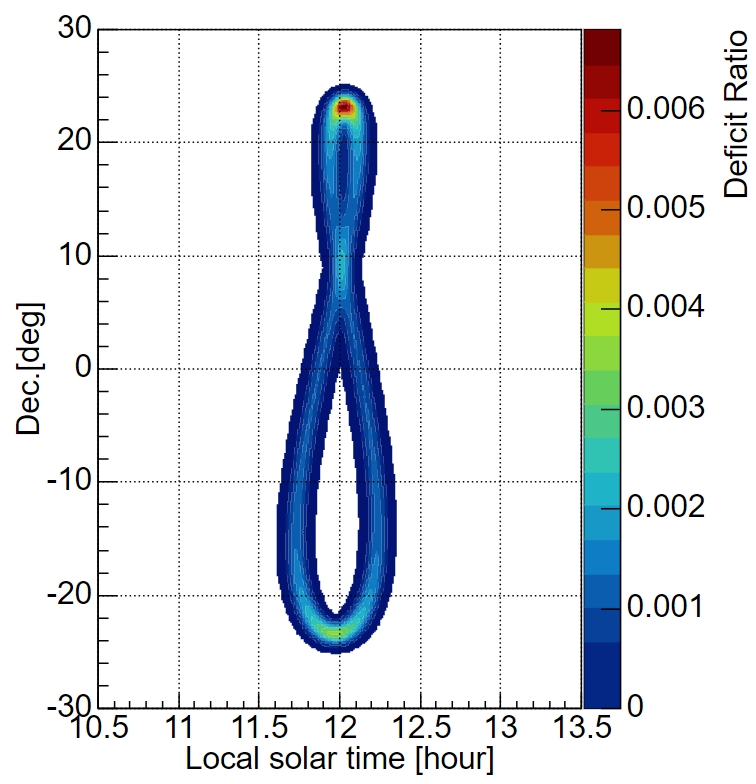}
    \caption{Two-dimensional deficit ratio map of the Sun's shadow in local solar time.}
    \label{fig:sol_2d}
\end{figure}

Fig. \ref{fig:sol_2d} displays a 2D map of the deficit ratio for the Sun's shadow, as observed at a site with a geographic latitude of \(0^\circ\). Around the 12th hour in local solar time, the Sun’s path forms an asymmetrical "8" shape, known as an analemma, due to the difference between mean and true local solar time \cite{Astro-book}. At higher declinations, the Sun remains longer in the same sky region, which increases the observed deficit ratio. This pattern reflects the combined effects of Earth's axial tilt and elliptical orbit, influencing the Sun's apparent motion across the sky.

\begin{figure}
    \centering
    \includegraphics[width=1\textwidth]{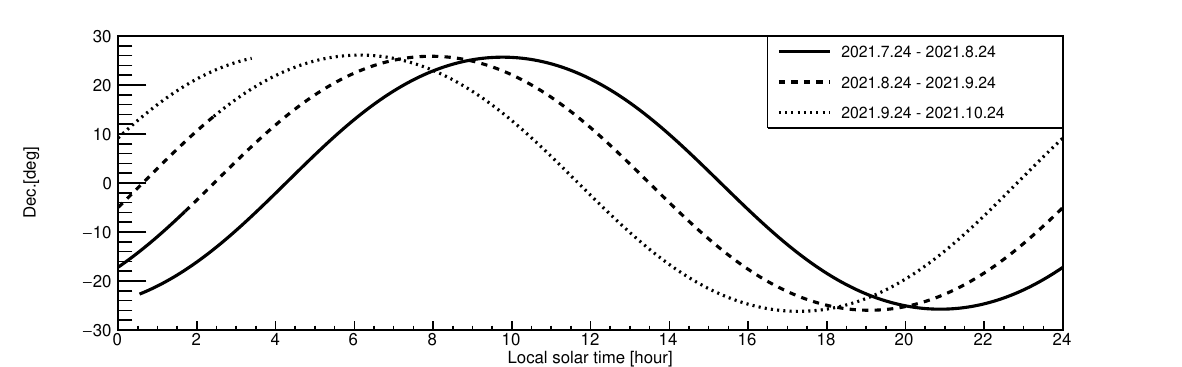}
    \caption{Example of the Moon's trajectory in the local solar time sky map from July 24, 2021, to October 24, 2021.}
    \label{fig:sol_moon_orbit}
\end{figure}

In contrast, the Moon's shadow does not produce a significant solar-time deficit due to its variable trajectory (see Fig. \ref{fig:sol_moon_orbit}). When a shadow occupies a single sky region (or grid cell) for an extended period, it blocks more cosmic rays relative to the background, thereby increasing the deficit ratio. However, the Moon’s path shifts daily and does not overlap in the same position each solar day, preventing its deficit from accumulating in the same manner. This variability in the Moon's trajectory results in its contribution to the total solar-time deficit being negligible. The Moon's constantly changing position prevents the buildup of a consistent shadow effect in any particular region of the sky, unlike the more stable path of the Sun.

\begin{figure}
    \centering
    \includegraphics[width=0.7\textwidth]{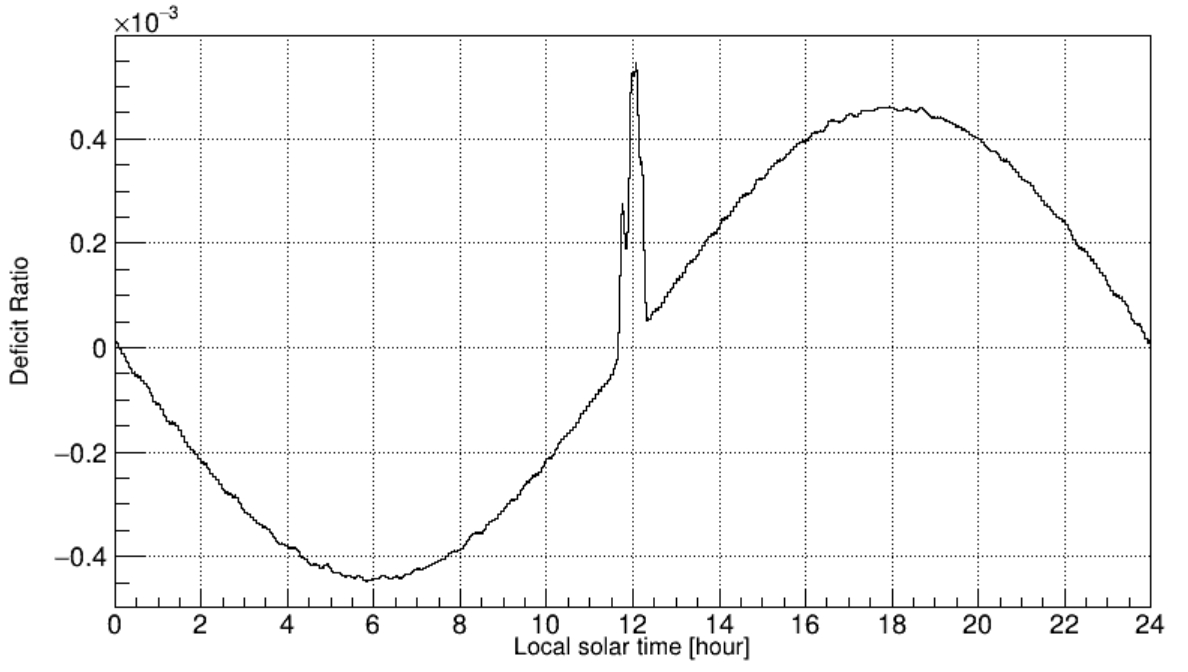}
    \caption{One-dimensional projection of the deficit ratio of the expected cosmic-ray anisotropy in local solar time. The peak at \(180^\circ\) is caused by the Sun's shadow.}
    \label{fig:cg_sol}
\end{figure}

Fig. \ref{fig:cg_sol} presents a 1D projection of the deficit ratio along right ascension in local solar time over a declination range of \(-50^\circ\) to \(50^\circ\), incorporating the Compton-Getting anisotropy and the effects of the Sun's and Moon's shadows. The cosine-like modulation represents the Compton-Getting effect, while the prominent deficit at the 12th hour in local solar time is attributed to the Sun's shadow. The irregularities and lack of smoothness in the curve are primarily due to the coupling of the Moon's shadow with the Compton-Getting effect, although the Moon's overall influence remains minor. This indicates that while the Sun's shadow contributes notably to the observed anisotropy, the Moon's variable trajectory introduces some fluctuations without substantially altering the overall pattern.

\begin{figure}
    \centering
    \includegraphics[width=0.7\textwidth]{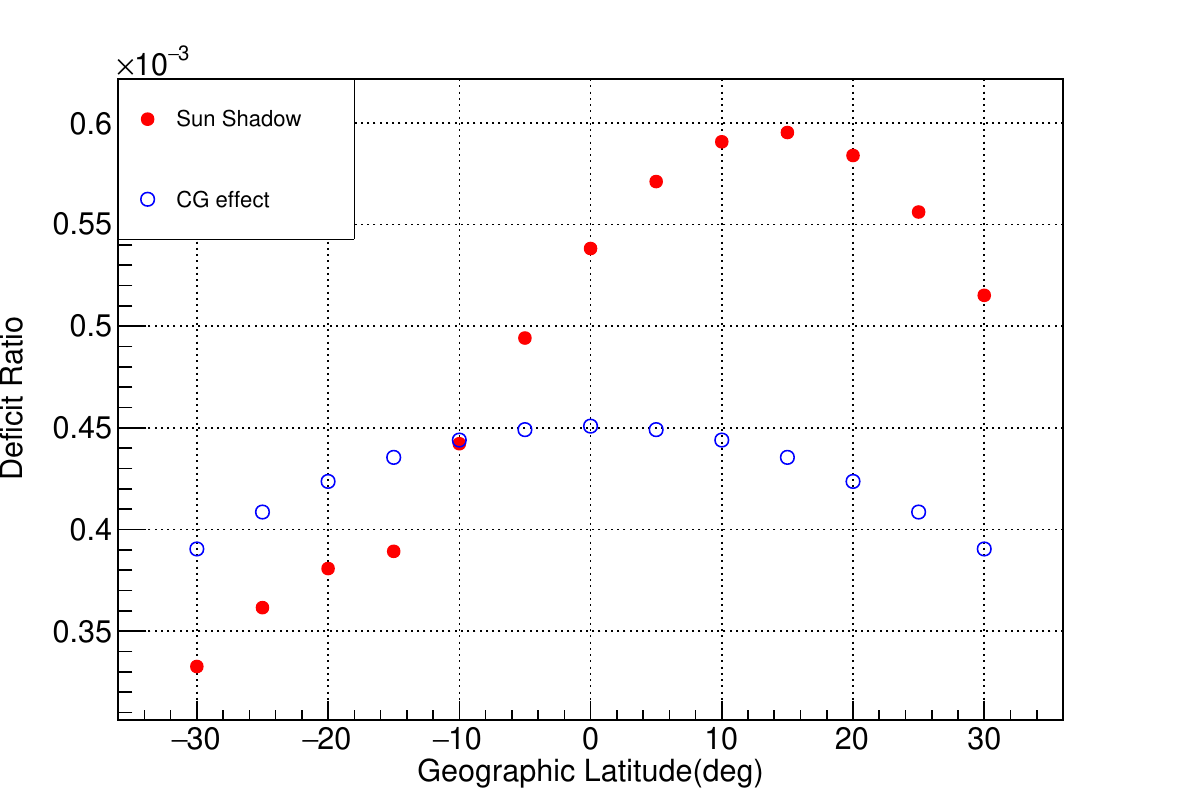}

    \caption{Maximum projected deficit ratio of the Compton-Getting anisotropy and the Sun's shadow effect as a function of the detector's geographic latitude in local solar time.}
    \label{fig:comp}
\end{figure}

The Compton-Getting effect and the Sun's shadow maintain similar magnitudes when the detector is at different geographic latitudes (see Fig. \ref{fig:comp}). Due to the Sun's path, the maximum deficit induced by the Sun's shadow remains around \(0.055\%\), while the Compton-Getting effect shows a slightly smaller maximum deficit of approximately \(0.045\%\). Regardless of the detector's location, the Sun's shadow produces a deficit comparable to that of the Compton-Getting effect.

\section{Conclusion}
\label{label:conclusion}
This study simulated the Sun's and Moon's shadows to analyze their impacts on cosmic-ray anisotropy in both local sidereal and solar time. In local sidereal time, the 1D projection along the right ascension of anisotropy due to the Sun's and Moon's shadows is approximately \(0.003\%\), which is negligible compared to the observed cosmic-ray anisotropy amplitude of around \(0.1\%\). However, in local solar time, the deficit ratio caused by the Sun's shadow is about \(0.055\%\), comparable to the amplitude of the Compton-Getting effect resulting from Earth's orbital motion. Consequently, in the case of local solar time, the influence of the Sun's shadow cannot be ignored and must be accurately estimated and removed from solar anisotropy data to ensure precise measurement and interpretation of cosmic-ray anisotropy.

The results in local solar time are sensitive to binning effects. In the simulation, the Sun's shadow deficit becomes significant at a bin size of \(0.1^\circ \times 0.1^\circ\); however, in actual observations, bins are typically larger, sometimes spanning several degrees. With larger bin sizes, the expected background counts of cosmic-ray events in a given bin increase, while the expected deficit event counts remain unchanged, making the Sun's shadow effect less noticeable. Consequently, the choice of bin size is crucial in accurately detecting and analyzing the influence of the Sun's shadow on cosmic-ray anisotropy in local solar time. Selecting an appropriate bin size ensures that the subtle effects of the Sun's shadow are not obscured by the larger background, allowing for a more precise assessment of its impact.

As statistical precision improves, the Sun's shadow effect becomes increasingly observable in experimental data. Future high-precision measurements are expected to reveal this effect more clearly. Moreover, some experiments have identified additional effects beyond the Compton-Getting effect \cite{ASgammaTeV}. The anticipated Sun's shadow effect could serve as a benchmark for validating the accuracy of these experimental observations.

\section*{Acknowledgements}
This work is supported by the National Natural Science Foundation of China (Nos. 12273114), the Project for Young Scientists in Basic Research of the Chinese Academy of Sciences (No.YSBR-061), and the Program for Innovative Talents and Entrepreneur in Jiangsu.




\bibliographystyle{apsrev}
\bibliography{refs}




\end{document}